\documentclass[electronic]{vgtc}             

\ifpdf
  \pdfoutput=1\relax                   
  \usepackage{graphicx}                
  \DeclareGraphicsExtensions{.pdf,.png,.jpg,.jpeg} 
\else
  \usepackage{graphicx}                
  \DeclareGraphicsExtensions{.eps}     
\fi%

\graphicspath{{figures/}{pictures/}{images/}{./}} 
\usepackage{tcolorbox}
\usepackage{microtype}                 
\PassOptionsToPackage{warn}{textcomp}  
\usepackage{textcomp}                  
\usepackage{mathptmx}                  
\usepackage{times}                     
\usepackage{cite}                      
\usepackage{tabu}                      
\usepackage{booktabs}                  
\usepackage{balance}

\onlineid{1471}

\vgtccategory{Main Research Track}

\vgtcinsertpkg

\title{{Using Virtual Reality for Detection and Intervention of \\Depression - A Systematic Literature Review}}

\author{Mohammad Waqas\thanks{e-mail: mohammad.wani@research.iiit.ac.in} %
\and Y Pawankumar Gururaj\thanks{e-mail: pawankumar.yendigeri@research.iiit.ac.in} %
\and V D  Shanmukha Mitra\thanks{e-mail: dhiraj.shanmukha@research.iiit.ac.in}    %
\and Sai Anirudh Karri\thanks{e-mail: saianirudh.karri@research.iiit.ac.in} %
\and Raghu Reddy\thanks{e-mail: raghu.reddy@iiit.ac.in}%
\and Syed Azeemuddin\thanks{e-mail: syed@iiit.ac.in} 
} %
\affiliation{\scriptsize  International Institute of Information Technology, Hyderabad , India}

\abstract{ 
The use of emerging technologies like Virtual Reality (VR) in therapeutic settings has increased in the past few years. By incorporating VR, a mental health condition like depression can be assessed effectively, while also providing personalized motivation and meaningful engagement for treatment purposes. The integration of external sensors further enhances the engagement of the subjects with the VR scenes. This paper presents a comprehensive review of existing literature on the detection and treatment of depression using VR. It explores various types of VR scenes, external hardware, innovative metrics, and targeted user studies conducted by researchers and professionals in the field. The paper also discusses potential requirements for designing VR scenes specifically tailored for depression assessment and treatment, with the aim of guiding future practitioners in this area.
} 

\CCScatlist{
  \CCScatTwelve{Human-centered computing}{Human computer interaction (HCI)}{Interaction devices}{Haptic devices};
  \CCScatTwelve{Hardware}{Communication hardware}{interfaces and storage}{Sensors and actuators}
  }

\begin{document}


\firstsection{Introduction}

\maketitle
According to the latest World Mental Health Report by the United Nations \cite{un}, there is a significant global burden of mental disorders, affecting nearly one billion individuals worldwide. This alarming statistic should be a matter of concern for people all around the globe. In recent times, the COVID-19 pandemic\footnote{https://www.who.int/news/item/02-03-2022-covid-19-pandemic-triggers-25-increase-in-prevalence-of-anxiety-and-depression-worldwide} has further exacerbated the mental health crisis. Factors such as job losses, the loss of loved ones, and various other stressors have contributed to an increase in mental health issues. Among these disorders, \textbf{Depression} has emerged as the most prevalent mental health problem, with a 25\% rise in its prevalence during the pandemic, as reported by the World Health Organization (WHO) \cite{un}.
 
\begin{tcolorbox}
\textbf{Depression} - \textit{a serious
mood disorder that causes a persistent feeling of sadness and
loss of interest} \cite{undepression}.
\end{tcolorbox}
The utilization of technology may have played a role in the rise of depression during the Pandemic \cite{t2d}. However, it is important to acknowledge that technological advancements and innovations have also been harnessed to identify and address these mental health disorders. Below are few approaches extracted from existing studies \cite{p63} \cite{q30}: 
\begin{itemize}
\item \textbf{Online counseling/support} - refers to a contemporary method of engaging with a psychotherapist or individuals facing similar challenges through the internet. This approach utilizes interactive voice technology and involves the utilization of disorder-specific questionnaires to facilitate the therapeutic process.  
\item \textbf{Short Messaging} - A commonly employed strategy for providing online reminders and brief messages as a means of motivating individuals to complete regular tasks according to specific schedules, thereby preventing personal crises. 
\item \textbf{Online Health Communities} -  
Typically, in the initial stages, focus groups often utilize these communities as a platform for exchanging daily routines and sharing experiences, thereby providing support and aiding in mood management.
\item \textbf{Wearable Devices} - Wearable devices are used in a specific situations where they need to monitor health parameters of the individual on a continuous basis.  For example, monitoring vital signs during physical activity to prevent personal collapse and provide immediate enhanced assistance. 
\item \textbf{Virtual Reality (VR)} - One of the emerging technologies being used for the detection and treatment of severe disorders by providing immersive experiences within a virtual environment.
\end{itemize}
The number of patients seeking mental health treatment has significantly increased, surpassing the number of trained medical professionals available. Traditional approaches to evaluating mental health, such as relying on psychotherapists using the DSM V, are subjective and prone to bias due to their reliance on patient recall. However, emerging technologies like Virtual Reality (VR) have the potential to greatly assist mental health practitioners in expediting the process of detecting and intervening in cases of depression. By providing ecologically valid environments \cite{q23} tailored to each individual, VR can aid in accurately identifying mental health symptoms. Consequently, VR-based assessment methods can assist medical practitioners in detecting the presence of depressive symptoms at an early stage, enabling timely medical interventions and improving the chances of a successful recovery. Following are the major contributions of this paper:
\begin{itemize}
    \item A compilation of VR scenes available in the existing literature used to examine the impact of depression. The scenes are summarized in Table \ref{tab:review}.
    \item Illustration of various physiological and behavioral response measures used to study depression along with the hardware used to monitor them. This is outlined in Fig \ref{fig:teaser}.
    \item The classification of review perspectives into various categories, including "Comparison among available conventional methods," "Meta Studies," and "Quantifiability and Dynamism," is conducted based on literature to assess the efficacy of virtual reality (VR) in identifying and intervening in depression.
    \item Recommendations to be considered by future practitioners for designing  VR scenes while studying depression. 
\end{itemize}

Rest of the paper is structured as follows: Section \ref{two} details our study methodology, research questions, search strategy, search string, and its search quality assessment. Section \ref{three} provides the search results and filtration process. Observations from the extracted literature are discussed in section \ref{four}. Section \ref{five} gives suggestions for designing a VR scene for depression. Section \ref{six} discusses the threats to the validity and   provides some conclusions.
\section{Study Setup} \label{two}
The study adheres to the systematic literature review protocols put forth by Kitchenham \cite{KitchenhamBBTBL09}. To provide a comprehensive understanding of the review study, the PICOC (Population, Intervention, Comparison, Outcome, and Context) method is employed to depict the context and significance. The Kitchenham approach is utilized to formulate the search string, search protocol, and research questions. The application of the PICOC method within the scope of this study is presented in Table \ref{picoc}. 

\begin{table}[ht]
\centering
\caption{PICOC of our Systematic Literature Review}
\begin{tabular}{|l|p{6cm}|} \hline
\label{picoc}
\textbf{Criteria}     & \textbf{Description}                            \\  \hline
\textit{Population}   & Researchers working on the identification and treatment of depression through the utilization of VR Scenes, as well as individuals passionate about developing software and hardware solutions that enhance both immersive and non-immersive VR devices. \\ \hline
\textit{Intervention} & Software and Hardware solutions complementary to VR HMD                                                      \\ \hline
\textit{Comparison}   & Different techniques based on VR scenes and hardware requirements   \\  \hline
\textit{Outcome}      & Effectiveness of VR in dealing with depression                                             \\ \hline
\textit{Context}      & Mental Health Community, VR Community of Academia       \\  \hline                                              
\end{tabular}
\end{table}

\subsection{Research Questions}
As stated in the previous section, the primary objective of the study is to detail the current research work in VR related to the detection and intervention of depression and evaluate its effectiveness. Below research questions are articulated to help us achieve our objective:
\begin{itemize}
    \item \textbf{RQ1} - \textit{What VR Scenes are used for the detection and intervention of Depression?}
    \item \textbf{RQ2} - \textit{What are the external hardware used as a complement to immersive and non immersive VR devices during detection and intervention of Depression?}
    \item \textbf{RQ3} - \textit{Are VR based methods of detection and intervention of Depression effective?}
\end{itemize}

\subsection{Search Strategy}\label{AA}
The search strategy was developed by taking into account the previous systematic literature reviews conducted in the healthcare field. Drawing upon the established practices within the healthcare domain, our research questions were formulated. Subsequently, a compilation of keywords that aligned with the research inquiries was assembled. To expand the scope of the search outcomes, the search string was refined by incorporating relevant synonyms. The search string was divided into two parts \textbf{S1} and \textbf{S2}
 \begin{tcolorbox}
     \textit{\textbf{S1:} ``VR'' OR ``Virtual Reality'' OR ``Virtual Environment''}
 \end{tcolorbox}
 \begin{tcolorbox}
     \textit{\textbf{S2:} ``Depression''}
 \end{tcolorbox}
The final formulated search string is \textbf{S1} AND \textbf{S2}.  The scope of search statement \textbf{S1} and \textbf{S2} is restricted to the \textit{``abstract''} rather than just the title to get the relevant papers. The final search string was the result of multiple iterations with peer researchers about the search keywords and the combination of the search keywords. 

\subsection{Inclusion and Exclusion Criteria}
The existing literature available in digital libraries such as ACM, IEEE Xplore, Pubmed, and Science Direct were the sources of study. From these libraries, we identified and gathered the pertinent papers that aligned with our search strategy and met the criteria for inclusion and exclusion.
\\
\newline
\textbf{Inclusion Criteria} -  This study exclusively focuses on research papers written in English. The timeframe for the papers considered spans from January 2000 to December 2022. In light of the emergence of VR in the past couple of decades, a deliberate choice was made to include only papers published after the year 2000. The scope of the study encompasses papers that explore both immersive and non-immersive VR applications for detection and intervention. Furthermore, papers from conferences, conference proceedings, and journals are all incorporated in the analysis.
\\
\newline
\textbf{Exclusion Criteria} -  
Papers lacking full-text availability are disregarded during the review process. Research contributions in the form of magazine articles, review notes, books, book chapters, and archives are also excluded from the review due to their insufficiency in meeting our research criteria. 

\subsection{Search Quality Assessment}
A set of seven questions were designed to further narrow down the research papers based on their relevance and reliability. This questionnaire awards a \textbf{yes}, \textbf{no} i.e. \textbf{1} or \textbf{0} where yes represents review consideration and no is the opposite. For a given paper it requires a summed score of 4 among the seven questions for review consideration. The questions are as follows
\begin{itemize}
\item Does the paper have a user study of value for further research?
\item Does the paper make a judgment on VR being better, worse, or the same as conventional methods?
\item Can we derive VR metrics from the paper?
\item Does the paper contain images of VR scenes?
\item Does the paper contain details of the VR scenes?
\item Are there external hardware complementing the immersive and non immersive VR devices?
\item Is there mention of findings, limitations, future scope, or discussions in the paper?
\end{itemize}

\section{Results} \label{three}
\begin{figure}[t]
     \centering
     \includegraphics[width=\columnwidth]{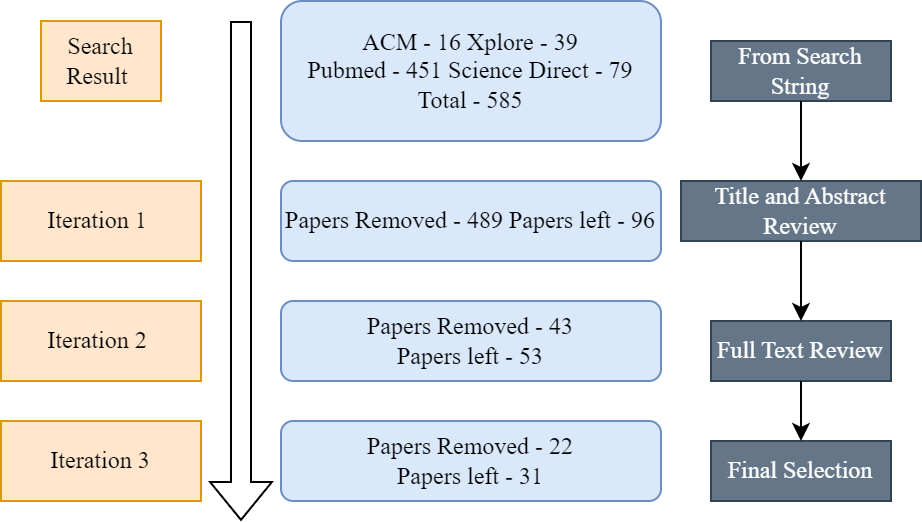}
     \caption{Paper Finalization after Search Strategy}
     \label{count}
    \end{figure}

Based on our initial search string, we extracted a total of  \textbf{\textit{584}} papers:  ACM – \textbf{\textit{16}}, IEEE Xplore – \textbf{\textit{39}}, Pubmed – \textbf{\textit{451}}, and Science Direct – \textbf{\textit{79}}.  Pubmed being a search engine for medical database, only systematic reviews were chosen because redundant papers and a  vast array of information irrelevant to this study had to be excluded. As a part of the first iteration, we removed duplicates, then applied inclusion and exclusion criteria to  further filter the paper set. Fig \ref{count} provides a detailed count of papers filtered as part of every iteration applied during our review process. Overall, \textbf{\textit{31}} research papers are considered for our study.
\section{Discussion} \label{four}
In this section, we discuss our observations with regard to the research questions
\begin{tcolorbox}
\textbf{RQ1: What VR scenes are used for the detection and intervention of Depression?}
\end{tcolorbox}
In the healthcare sector, the use of virtual reality scenes via a Head Mounted Device is described to be immersive in nature. The Virtual Environment is the primary stimulus that provides the user the experience of a customized, immersive, and engaging environment.  
In the VR Scenes, we have included scenes as well as games because games inherently comprise a sequence of scenes. For example, in a simulation of a refreshing hike through the woods, the user, through neurofeedback using Electroencephalography controls the sunlight and bird-singing sound in the gaming environment. Another scene is a dolphin swimming in the deep ocean. Neurofeedback reflects the speed at which the dolphin swims and the school of fish it encounters on its way\cite{q1}. This is used an intervention mechanism for depression where the person after playing the game after 10 minutes feels relaxed.

In \cite{q13} the VR scene mimics the ambience of a waiting room in a VR mental health clinic with doctor avatars and boards, where the participants have to complete the questionnaires on the boards and spot the difference between two identical rooms. Eye tracker is used to collect data like the gaze origin, gaze direction, the location where the eyes converge, and the confidence value. Metacognitive skills for symptom identification based on how accurately the user performed the task are used as a measure for the detection of depression.

In \cite{q6}, the VR scene tries to simulate a psychologically adverse environment using a negative text message in a dark background. The participant's job was to make the message disappear using a variety of means like crumbling it, punching it or stabbing it using a lance of light. The user can also change the message to a positive one. It is an intervention mechanism to help people cope with negative thoughts and emotions.

Table \ref{table:VRsceneDetectInterveneDepression} provides the scene description, task the participant has to perform, whether the VR scene is immersive or non-immersive, Metrics recorded, disorder relating to the scene, and whether it is used for detection or intervention.
\begin{tcolorbox}
\textbf{RQ2:  What are the external hardware used as a complement to immersive and non immersive VR devices  during detection and intervention of Depression?}
\end{tcolorbox}

\begin{figure*}[t]
     \centering
     \includegraphics[scale= 0.250]{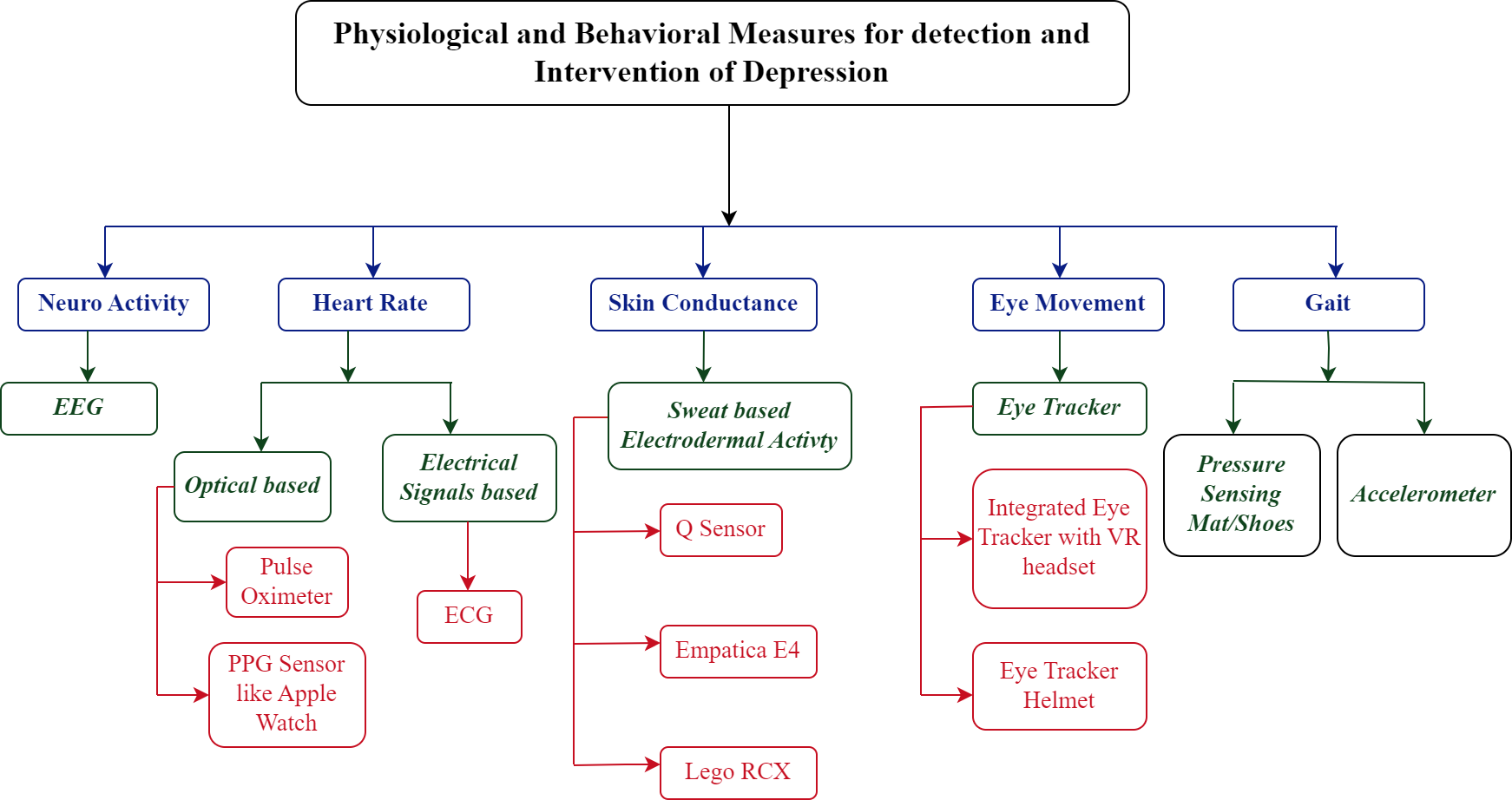}
     \caption{Categorization of the Hardware used in detection and intervention of depression}
     \label{fig:teaser}
    \end{figure*}

Before embarking on our discourse about the external hardware, it is imperative to briefly explain immersive and non immersive VR. In immersive, a Head Mounted Device(HMD) or CAVE gets used as a display, and in non immersive, the environment is displayed on a screen. 
Talking about non immersive VR, the scene is displayed on a laptop as in \cite{q8}, PCs and projectors as in \cite{q18}\cite{q9}\cite{q27}. In \cite{q11}, both immersive and non-immersive forms of VR were used, whereas in non immersive it used a laptop.
\\
\newline
\textbf{Physiological and Behavioural Measures}: Along with the HMDs or the PCs other hardware is present to record the physiological and behavioral response of the participant to the VR scenes shown.
The physiological and behavioural measures shown to have a have correlation with depression or emotional state of a person are brain wave activity or \textbf{Neuro activity}  \cite{q2}, \textbf{heart rate variability} \cite{q35}, \textbf{skin conduction} \cite{skincon_dep}, \textbf{eye movement/pupil waves}\cite{q38}\cite{q8} and \textbf{gait} \cite{gait_dep}.
The sensors used to measure these parameters are mentioned in Fig \ref{fig:teaser}. Information pertaining to the sensors is laid down below.

Electroencephalography (EEG) stands as an electrophysiological monitoring technique, facilitating the capture of the brain's electrical activity. This non-invasive method has found utility in the examination of cognitive behaviors, particularly the intricate patterns of brain wave activity closely associated with conditions such as depression \cite{EEG_dep}. Within the research documented in \cite{q1}, a three-electrode EEG collector, distinguished for its correlation with emotional processes, was employed as a cost-effective alternative to the more elaborate 64 or 128 electrode systems.

Furthermore, in the study detailed in \cite{q2}, a pair of EEG setups was utilized: one featuring seven sensors and the other employing five sensors. Complementing these EEG recordings, continuous heart rate data was collected and stored through the use of an Apple Watch. This monitoring was rooted in photoplethysmography, an ingenious technique incorporating LED lights and photodiode sensors discreetly positioned on the watch's underside. This approach effectively measures data from the cardiovascular system, notably showcasing its reliability in assessing heart rate, particularly in scenarios involving anxiety. Impressively, the sensor has a wide operational range, spanning from 30 to 210 beats per minute while requiring minimal active participation from the study participants.

In addition, the study described in \cite{q2} also harnessed the capabilities of a Q sensor for the measurement of skin conductance. Affixed to the wrist, this device incorporates a small sensor at its base, enabling the recording of Electrodermal Activity (EDA) in the skin. The observed fluctuations in EDA are subsequently relayed to specialized software (Q-live). The underlying principle of this technology lies in its ability to employ electrical signals of a remarkably low power, measuring less than five micro watts, to gauge skin conductance.

In the research outlined in \cite{q5}, an oximeter served as a valuable instrument for measuring both heart rate variability and peripheral capillary oxygen saturation (SPO2) levels. These metrics, in turn, were leveraged to assess the emotional state of the participants. Notably, a body of research has underscored the intricate connection between heart rate and various forms of stress, where heightened stress levels have been associated with the onset of depression \cite{q35}. Furthermore, heart rate variability has emerged as a useful indicator for gauging psychological well-being and the presence of mental stress \cite{q36}.

In another research a Galvanic Skin Response sensor(GSR) which senses changes in sweat gland activity that accurately show a person's emotional state was used \cite{q37}. The two metrics for the  GSR are skin conductance and skin resistance. It is attached to the fingers on the index and the middle fingers where the current is sent from one and the other measures the difference. The readings are stored and displayed in a graph where peaks are highlighted and values are smoothed by averaging them..

Li, Mi, et al. in \cite{q8}, looked at pupil diameter employing specialized eye movement helmets for data collection. Unlike behavioral signals like facial expressions and speech, pupil diameter offers a direct and unfiltered insight into emotional expression as it increases on a joyful emotional expression and decreases on a sad emotional experience.\cite{q38}. The researchers called this dynamic physiological signal as pupil wave. This unique attribute positions them as an objective and reliable signal for assessing an individual's mental state. Furthermore, when coupled with emotion-inducing virtual reality (VR) scenarios, the use of pupil waves enhances the authenticity of the emotional experience.


In the user study in \cite{q9}, The skin resistance signal acquisition is done by LEGO RCX Mindstorm. The working principle is that a constant voltage is applied to the skin by the electrodes and the current that passes over the skin is detected and displayed.
The constant voltage is applied by  the GSR amplifier to the skin by electrodes. The voltage is very small that it cannot be perceived. The current that passes over the skin, as the voltage is applied, can be detected and displayed. Silver(Ag) plates are used to construct the electrodes and Spectra 360 electrode gel is applied for better skin contact. Readings are sent to the GSR interface that converts them to a graph.

In \cite{q11} the Neuroactivy using EEG signals was collected using a Muse headband with 4 channels (TP9, FP1, FP2, TP10) and the sampling rate was 256 Hz.

In \cite{q13}
 a VR headset outfitted with an eyetracker within the mark was used as Eye-tracking sensing technology which is used to collect non-self-report data from the VR simulations to understand abnormalities in eye movements correlated with mental health disorders. The eye-tracking data includes: (1) Gaze Origin: the origin of a ray cast from the eye, (2) Gaze Direction: a 3D vector that shows what direction the wearer is looking, (3) Fixation point: the location of where the eyes converge, (4) and Confidence Value: a number [0, 1] that denotes the reliability of the sensor

The study \cite{q15} involved recording Neuroactivty using EEG from 64 electrodes positioned according to the international 10–20 system (AF3/4, AF7/8, FPz, Fz, Fp1/2, F1/2, F3/4, F5/6, F7/8, FCz, FC1/2, FC3/4, FC5/6, FT7/8, Cz, C1/2, C3/4, C5/6, T3/4, T7/8, TP7/8, CP1/2, CP3/4, CP5/6, Pz, P3/4, P5/6, P7/8, PO3/4, PO5/6 PO7/8, O1/2, POz, Oz, HEOR, HEOL, ECG, VEOU, VEOL) and referenced to CPz. For data recording the EEG amplifier produced by Neuroscan was used at 1000 Hz sampling rate. The impedance of all electrodes dropped below 10 k$\mathrm{\Omega}$ during the experiment. 60 Hz online bandpass filter and 50 Hz notch filter were used to filter high frequency noise, baseline drift, and power line interference.

 Empatica E4, a research-grade wristband wearable was used in \cite{q20} where its two silver-plated electrodes measure electrodermal activity (i.e. Skin Conductance) at the inner wrist at a 4 Hz sampling rate; Heart Rate was determined using a photoplethysmography (PPG) sensor, which is based on the principle of more blood oxygenation means that more light is absorbed so it measures the Blood Volume Pulse(BVP) using green and red LEDs emitting light that is reflected as a function of blood oxygenation. During a heartbeat, less light is therefore reflected. The BVP signal is measured at a sampling frequency of 64 Hz. From this BVP signal, HR was calculated. The Gait i.e. the walking pattern or movement was recorded using a 3 axis accelerometer. A pressure sensing mat can also be used for recording gait\cite{q39}. The session was also videotaped using a handheld camera stationed on a tripod.
 Finally in \cite{q22} ECG signals i.e. where the heart's electrical activity is tracked were also recorded along with the Neuroactivty using EEG signals.
\newline In Summary, we categorized the physiological and behavioral measures as well as the associated sensors as shown in Figure \ref{fig:teaser}.
\begin{table*}[t]
\centering
\caption{Illustration of Scenes in VR for Detection or Intervention on Depression Studies}
\label{table:VRsceneDetectInterveneDepression}
\resizebox{\textwidth}{!}{%
\begin{tabular}
{|p{4cm}|p{4cm}|p{2cm}|p{4cm}|p{2.5cm}|p{1cm}|p{0.5cm}|p{0.25cm}|}
\hline
\textbf{Scene description} &
  \multicolumn{1}{c|}{\textbf{Task to be Performed}} &
  \multicolumn{1}{c|}{\textbf{Type of VR}} &
  \multicolumn{1}{c|}{\textbf{Metrics Recorded}} &
  \multicolumn{1}{c|}{\textbf{Disorder}} &
  \multicolumn{1}{c|}{\textbf{Detection/Intervention}} &
  \multicolumn{1}{c|}{\textbf{Year}} &
  \multicolumn{1}{c|}{\textbf{Reference}} \\ \hline

  A Hike Through the woods and a dolphin swimming &
  Head Movement : The neuro-feedback from the head will control the  sunlight in wood and the speed in Dolphin &
  Immersive HMD based &
  Neuroactivity  &
  Depression &
  Intervention &
  2017 &
  \cite{q1} \\ \hline

  Top view of a city with large buildings &
  Stand on a plank on top of a building and look down &
  Immersive HMD based &
  Neuroactivty, Heart Rate, Skin Conductance &
  Anxiety &
  Intervention &
  2019 &
  \cite{q2} \\ \hline

  Middle class house with stressors like newborn baby crying, telephone ringing, fire in the kitchen &
  Explore the combination of two or more stressors &
  Immersive Virtual Glass based &
  No &
  Post Natal Depression &
  Intervention &
  2020 &
  \cite{q3} \\ \hline

  Natural scenes like mountains, island and bodies of water categorized in seasonal scenes viz summer, winter, autumn, spring &
  Explore the environment for as long as possible &
  Immersive HMD based &
  No &
  Depression &
  Intervention &
  2020 &
  \cite{q4} \\ \hline

  Places of Philippines having natural beauty &
  Explore the environment &
  Immersive HMD based &
  Heart Rate, Skin Conductance &
  Stress &
  Intervention &
  2020 &
  \cite{q5} \\ \hline

  A Negative text message in a dark background &
  Change the message into a positive one or crumble it, punch it or stab the message to make it disappear &
  Immersive Cardboard based VR headset &
  No &
  Depression &
  Intervention &
  2021 &
  \cite{q6} \\ \hline

  Customized Avatar &
  Customize the eyes, hair, gender as well as the environment as a room, beach, castle and interact with the avatar &
  Immersive HMD based &
  No &
  Depression &
  Intervention &
  2021 &
  \cite{q7} \\ \hline

  Different environments that elicit sadness, calm or joy &
  Explore the Environment &
  Non immersive Laptop based &
  Pupil Waves &
  Depression &
  Detection &
  2022 &
  \cite{q8}\\ \hline

  A Virtual bar and a beach &
  Explore the environment &
  Non immersive PC based &
  Skin Conductance &
  Alcoholic Depression &
  Intervention &
  2013 &
  \cite{q9} \\ \hline

  Forest and a shelter tower in a grass field with flowers &
  Explore while growing sunflowers showed office work like emails &
  Immersive HMD based &
  No &
  Stress &
  Intervention &
  2017 &
  \cite{q10} \\ \hline

  Nature with trees and grass and roller coaster ride &
  Explore the Environment &
  Immersive HMD based as well as non immersive laptop based &
  Neuroactivity &
  Depression, Anxiety, Psychosis &
  Detection,Intervention &
  2019 &
  \cite{q11} \\ \hline

  Childhood Avatar of the Participant &
  Interact with the sad, happy, and fearful avatar &
  Immersive HMD based &
  No &
  Depression, Anxiety &
  Intervention &
  2020 &
 \cite{q12} \\ \hline

  Waiting room in the mental health clinic, Doctor avatars and boards &
  Perform the tasks as directed like completing questionnaires &
  Immersive HMD based &
  Gaze origin and direction, 
  fixation point &
  Depression &
  Detection &
  2020 &
  \cite{q13}\\ \hline

  A Dark dissolute scene followed by fire in the midst of darkness and peaceful weather &
  Explore and progress by collecting supplies including medication &
  Immersive HMD based &
  No &
  Depression &
  Intervention &
  2021 &
  \cite{q14} \\ \hline
  Multiple Spherical balls changing colour in a Sphere and returning back to the original colour &
  Choose the ones that changed colour &
  Immersive HMD based &
  Neuroactivty &
  Depression, Anxiety &
  Detection, Intervention &
  2022 &
  \cite{q15} \\ \hline

  Google VR Earth &
  Place oneself anywhere in the world and experience the environment &
  Immersive HMD based &
  No &
  Depression &
  Intervention &
  2021 &
  \cite{q16} \\ \hline

  A Virtual Lunchroom &
  Move through the room while being laughed at by other guests &
  Immersive HMD based &
  Skin Conductance, Heart Rate, Gait &
  Depression &
  Intervention &
  2021 &
  \cite{q20} \\ \hline
\end{tabular}%
}
\label{tab:review}
\end{table*}

\begin{tcolorbox}
\textbf{RQ3: Are VR based methods for detection and intervention of Depression effective?}
\end{tcolorbox}

As we progress into the future, to address the problems of expensive depression healthcare and the shortage of specialised healthcare workers, it is imperative that the efficacy of VR be established among the landscape of emerging technologies. For that, we have identified different metrics as follows:
\newline
\\
\textbf{User Study} - One of the most common approaches used to assess the effectiveness of Virtual Reality (VR) is through user studies. The reviewed papers had participants from diverse ages\cite{q7}, genders\cite{q1} and professions\cite{q6}. In one study \cite{q1}, 12 participants took part, with 11 of them reporting the effectiveness of VR and expressing a willingness to use it again.Similarly, in a study \cite{q3} involving six participants, VR scenes featuring stress-inducing elements were presented. Participants' responses were collected through questionnaires like the Edinburgh Postnatal Depression Scale (EPDS) and the Generalised Anxiety Disorder questionnaire (GAD-7). This preliminary investigation yielded positive outcomes, as all six participants reported feeling better, experiencing increased relaxation, improved mood, enhanced self-esteem, and better sleep and appetite following the therapy.
In another study \cite{q4}, mood enhancement was observed among 20 participants (comprising 10 males and 10 females) engaged in a user study. However, it is important to note that this positive effect was accompanied by side effects such as eye strain, nausea, and headaches. These side effects were generally tolerable for most participants, but it is worth mentioning that individuals with epilepsy may be more susceptible to experiencing adverse symptoms from VR use.

A substantial study involving 295 participants from various institutions (including Beijing Anding Hospital Capital Medical University, Beijing University of Technology, military regions 1 and 2) was conducted for depression detection and yielded lower detection errors compared to previous methods \cite{q8}.
In yet another study \cite{q16}, which focused on 16 females and two males diagnosed with moderate to moderately severe depression, VR intervention resulted in increased positive emotions and decreased negative emotions.
However, in a study involving 18 participants, documented in \cite{q2}, no statistically significant differences were found in heart rate increase or self-reported anxiety between baseline and VR recovery phases, regardless of whether participants had higher or lower anxiety scores.
\\
\newline
\textbf{Limitations: }In summary, while most studies have highlighted the potential of VR in depression detection and intervention, it is important to acknowledge that these studies are predominantly pilot studies with relatively small sample sizes. Barring few studies which reported eye strain, headache, dizziness there a little to no consideration for ethical issues as such potential psychological distress for the participants. The frequency, duration, and intensity of VR scenes using HMD that are safe for a diverse range (age, professional, gender) of populations need to be discussed. Further comprehensive investigations are warranted to generalize as well as solidify the effectiveness of VR in this domain. Table \ref{table:EffectivenessVR} illustrates some of the VR scenes based on the method used and the potential results observed.
\\
\newline
\textbf{Ability to Elicit Emotion} -  
Immersive Virtual Reality (VR) has been shown to elicit more intense emotional responses, thus positioning it as a promising tool for future emotional studies \cite{q11}. In a study involving 41 healthy students, detailed in \cite{q20}, findings from three physiological metrics—specifically, the frequency of Skin Conductance (SC) peaks per minute, average SC levels, and mean Heart Rate (HR)—revealed statistically significant increases in arousal levels when participants engaged with VR compared to the conventional, face-to-face imaginative experiences conducted with an experimenter.
Similarly, within the scope of VR driving simulations, a comprehensive study with a substantial group of 463 participants, as chronicled in \cite{q22}, substantiated the VR's capacity to evoke emotions and elicit corresponding physiological responses.
Moreover, the investigation presented in \cite{q33}, comprising 30 participants, established a noteworthy correlation between the emotional valence of the VR experience and the participants' spatial and temporal navigation within the virtual environment. Specifically, a positive and pleasant experience was found to facilitate more extensive exploration of the virtual space compared to encounters characterized by negative emotions.
\\
\newline
\textbf{Improved Detection and Intervention} - Within the context of the research detailed in \cite{q8}, the attempt was the identification of depressive states through the analysis of eye movements, which were meticulously recorded during interactions with emotionally charged Virtual Reality (VR) scenarios. The key here was the utilization of pupil-wave data, a metric capable of directly conveying the user's emotional state. To facilitate this assessment, a novel computational framework was devised, featuring a dual-channel one-dimensional convolutional neural network (CNN). This CNN architecture incorporated cascade parallel multi-scale convolutional residual blocks and width-channel attention modules, and was adeptly harnessed for both depression and anxiety level assessments. Notably, the results demonstrated a reduced margin of error in comparison to prior methodologies.
In a distinct investigation mentioned in \cite{q19}, a group of 29 older adults was engaged in a study focusing on the impact of VR interventions on happiness levels. The intervention group, which experienced VR interventions, exhibited a noteworthy increase in happiness compared to the control group, which received no such interventions.

\begin{table*}[ht]
\caption{Effectiveness of Virtual Reality on Observed Disorder type, methods and results}
\label{table:EffectivenessVR}
\label{tab:my-table2}
\resizebox{\textwidth}{!}{%
\begin{tabular}{|c|c|c|c|c|c|}
\hline
\textbf{Scene description} &
  \textbf{Type of VR} &
  \textbf{Metrics Recorded} &
  \textbf{Disorder} &
  \textbf{Methods} &
  \textbf{Results} \\ \hline
\begin{tabular}[c]{@{}c@{}}A Hike Through the woods and \\ a dolphin swimming\end{tabular} &
  Immersive HMD based &
  Neuroactivty &
  Depression &
  User Study &
  11 out of 12 people found it effective \\ \hline
Top view of a city with large buildings &
  Immersive HMD based &
  \begin{tabular}[c]{@{}c@{}}Neuroactivty, Heart Rate, \\ Skin Conductance\end{tabular} &
  Anxiety &
  User Study &
  \begin{tabular}[c]{@{}c@{}}Severe cases: No statistical difference\\ Moderate to less case: Increase in heart rate\end{tabular} \\ \hline
\begin{tabular}[c]{@{}c@{}}Middle class house with stressors like \\ newborn baby crying, telephone ringing, \\ fire in the kitchen\end{tabular} &
  Immersive Virtual glass based &
  Mood, self esteem, sleep &
  Post Natal Depression &
  User Study &
  \begin{tabular}[c]{@{}c@{}}All 6 users reported better sleep, \\ mood and self esteem\end{tabular} \\ \hline
\begin{tabular}[c]{@{}c@{}}Natural scenes like mountains, island and \\ bodies of water categorized in seasonal \\ scenes viz summer, winter, autumn, spring\end{tabular} &
  Immersive HMD based &
  Mood &
  Depression &
  User Study &
  \begin{tabular}[c]{@{}c@{}}Mood Improved but VR HMD \\ also caused eye strain, nausea \\ and headache\end{tabular} \\ \hline
A Negative text message in a dark background &
  \begin{tabular}[c]{@{}c@{}}Immersive Cardboard \\ based VR headset\end{tabular} &
  \begin{tabular}[c]{@{}c@{}}Subjective Experience \\ described through \\ Interviewing\end{tabular} &
  Depression &
  User Study &
  \begin{tabular}[c]{@{}c@{}}All 10 people had a positive \\ experience with VR\end{tabular} \\ \hline
\begin{tabular}[c]{@{}c@{}}Different environments that \\ elicit sadness, calm or joy\end{tabular} &
  Non immersive Laptop based &
  Pupil Waves &
  Depression &
  User Study &
  \begin{tabular}[c]{@{}c@{}}In the 295 samples, error in the \\ detection of depression was \\ less as compared to previous \\ methods\end{tabular} \\ \hline
A Virtual bar and a beach &
  Non immersive PC based &
  Skin Conductance &
  \begin{tabular}[c]{@{}c@{}}Alcoholic \\ Depression\end{tabular} &
  User Study &
  \begin{tabular}[c]{@{}c@{}}In the 5 people based on skin \\ conductance results VR was \\ effective in eliciting as well as \\ calming the craving\end{tabular} \\ \hline
\end{tabular}%
}
\end{table*}

\textbf{ Comparison with conventional Methods} - In the study documented in \cite{q8}, the collection and analysis of eye movement data provide a quantitative and testable way of assessing depression as compared to the conventional method(of visiting a psychiatrist and answering questions), where there is recall bias. Furthermore in \cite{q12}, the VR based method of joyful, sad, fearful childhood avatar performed better in the intervention of depression than the conventional method. This novel VR-based approach outperformed conventional methods, which involve merely looking at childhood photos and attempting to conjure emotional scenarios. The results indicated the superior efficacy of VR in the context of depression intervention. Lastly in \cite{q18} the cognitive behavioral program with VR was as effective as standard Cognitive Behavioural Therapy(CBT) for treating depression, and the statistically significant differences were in favour of VR.
\\
\newline
\textbf{Meta Studies} -
In a comprehensive examination detailed in \cite{q25}, encompassing a meta-analysis of 11 studies and 6 Randomized Controlled Trials (RCTs), the researchers arrived at a compelling consensus. They determined that Virtual Reality (VR) exhibited a distinct and favorable impact on patients grappling with depression. Furthermore, in an exhaustive analysis of 18 RCTs expounded upon in \cite{q26}, the research team scrutinized the effects of VR-based games incorporating bodily movement, particularly among the elderly demographic. The findings illuminated a positive influence on memory, cognitive abilities, and depression within this population. Nevertheless, it is noteworthy that the literature presents some divergent perspectives. For instance, the meta-analysis featured in \cite{q31} concluded that serious games held promise as a means to alleviate depression. Conversely, the investigation reported in \cite{q32} indicated that gaming failed to surpass conventional methodologies in this regard, emphasizing the necessity for more high-quality RCTs to establish a definitive stance.
\\
\newline
\textbf{Quantifiability and Dynamism} - In the research described in \cite{q3}, the stress-inducing elements embedded within the Virtual Reality (VR) environment exhibit a dynamic nature. These stressors can be modified by the therapist, who possesses the flexibility to alter their attributes, including their frequency, duration, and whether they are administered individually or concurrently. Measuring and recording Neuroactivity using EEG signals, heart rate, pupil waves, skin conductance and gait measured while the participants are under the influence of the immersive VR environment provides a quantitative measure. These readings can be then sent to an expert for further analysis. This quantitative approach proves superior to the subjective tendencies associated with memory recall bias. 
\\
\newline
\textbf{Negative User Feedback} - In the study presented in \cite{q10}, a total of 32 participants engaged in a Virtual Reality (VR) intervention. Notably, among these participants, 9 individuals, or approximately 28\%, reported experiencing feelings of dizziness. However, it is noteworthy that a significant majority, precisely 87.5\% of the participants, expressed a strong inclination toward utilizing VR as part of their future experiences. In a related investigation detailed in \cite{q24}, involving elderly individuals as subjects, a sample size of 5 participants was employed. Regrettably, the findings indicated that most of these elderly participants could not comfortably employ the VR headset for durations exceeding 5 minutes due to the onset of dizziness. Furthermore, in the context of the research documented in \cite{q4}, participants voiced complaints about various discomforts while engaging with VR. These included eye strain, nausea, and headaches. Such adverse effects are crucial considerations when implementing VR interventions.
\section{Future VR Scenes for Depression Studies} \label{five}
The Table \ref{tab:review} provides details of different VR scenes and their corresponding recorded physiological and behavioural measures. Figure \ref{fig:teaser} shows the observed sensors to measure these parameters. By thoroughly examining the observations from this review study, we formulated a bare-minimum checklist to design VR environments specifically tailored for studying depression with requisite sensors.
\begin{itemize}
    \item \textbf{Measures Selection:} Before embarking on the design of a VR scene, it is required to have a careful consideration of physiological and behavioral metrics that are required for assessment.
    \item \textbf{Content Impact: }The effectiveness of a VR scene depends upon its capacity to elicit pronounced responses in the measured physiological parameters. Hence, the scene's content should be thoughtfully designed to maximize its influence on these metrics.
    \item \textbf{Baseline Values: }To facilitate meaningful comparisons, it is required to establish baseline values for the metrics of interest prior to VR stimulation. In cases where these baselines are unknown, they should be calculated through pre-stimulation measurements
\end{itemize}
Let us consider an example to understand these principles in practice. Consider a study where we would like to measure  physiological reactions such as increased \textbf{heart rate} and \textbf{perspiration} for a group of individuals. For such an experiment setup, we can simulate a plank atop a tall structure in our VR Scene (Content Impact). Compelling participants will gaze downward thus meticulously crafting the focus on eliciting variability in heart rate and skin conductance. To ensure a robust comparison, baseline heart rate and skin conductance values is to be captured during participant comfort and leisure.

The versatility of VR scenes is exemplified by the concept of a virtual lunchroom, wherein participants navigate within the virtual space while subject to various levels of ridicule from virtual guests. This scenario can be customized by adjusting the number of virtual guests and their position in the scene to increase or decrease content impact while monitoring metrics like skin conductance, heart rate, and gait. The baseline values can then be compared with the different scenes with variable content impact.

Furthermore, it is noteworthy that VR scenes designed for therapeutic intervention can also be repurposed for detection. For instance, a VR scene featuring a customized avatar to foster self-compassion can be transformed into a detection tool. By integrating an eye-tracking system and allowing participants to select different emotional avatars (positive, negative, or calm), it is possible to record and assess pupil wave patterns for indications of depression levels, as demonstrated in the work of Li et al \cite{q8}.
\section{Threats to Validity} \label{six}
Our search strategy was extensive, and our observations are factual as per the reviewed research articles. Our genuine insights can be replicated by repeating the study using the same search strategy. We engaged with our peer reviewers and VR practitioners to assist us with search string finalization, filtration, review, and analysis. Throughout the review, we received constant feedback on our search strategy. The possibilities of a few primary studies being overlooked are limited. Of course, authors could make minor mistakes regarding the judgment of a research paper during the filtration process. The peer researchers agree on search Strings. We have attempted to conduct this review under a systematic review protocol. It is important to acknowledge that variations in results may arise if the search strategy and data extraction are conducted using a different protocol in the future.
\section{Conclusion} \label{seven}
The immersive nature of Virtual Reality (VR), coupled with its adaptability, cost-effectiveness, motivational potential, and ability to evoke emotions, positions it as a highly promising candidate for addressing depression when compared to alternative emerging technologies. Furthermore, it is imperative to employ the appropriate hardware to capture physiological and behavioural responses elicited by VR stimuli effectively. In this research endeavor, we have elucidated the specific VR scenes employed to stimulate responses for the detection and intervention of depression. Moreover, we have delineated a spectrum of physiological measures associated with depression, encompassing neuro-activity (assessed via EEG), heart rate, skin conductance, eye movement, and gait. To facilitate the selection of suitable hardware for measuring these physiological responses, we have devised a comprehensive categorization that aligns each measure with the requisite sensor. Furthermore, we have developed a structured checklist for the development of VR scenes tailored for depression intervention. This checklist is substantiated with concrete examples from existing research. These resources will empower VR practitioners engaged in depression-related research to make informed choices based on their unique requirements.In sum, we aspire for this review to serve as a pivotal step towards reducing challenges tied to subjective bias, the high cost of treatment, and the dearth of trained professionals in depression detection and intervention.
\balance
\bibliographystyle{abbrv-doi}

\bibliography{template}
\end{document}